\documentclass[prl,reprint,nofootinbib,showpacs,superscriptaddress,twocolumn]{revtex4-1}

\usepackage{graphicx}
\usepackage{hyperref}
\usepackage{amsfonts}
\usepackage{amsmath,amssymb}
\usepackage{bm}
\usepackage{color}

\usepackage{epsfig}

\def\be{\begin{equation}}
\def\ee{\end{equation}}
\def\bea{\begin{eqnarray}}
\def\eea{\end{eqnarray}}
\def\ba{\begin{array}}
\def\ea{\end{array}}
\def\bem{\begin{multline}}
\def\eem{\end{multline}}

\begin{document}

\title{Quantum machine learning over infinite dimensions}

\author{Hoi-Kwan Lau}
\affiliation{Institute of Theoretical Physics, Ulm University, Albert-Einstein-Allee 11, 89069 Ulm, Germany}
\author{Raphael Pooser}
\affiliation{Quantum Information Science Group, Oak Ridge National Laboratory, Oak Ridge, Tennessee 37831, U.S.A}
\affiliation{Department of Physics and Astronomy, The University of Tennessee, Knoxville, Tennessee 37996-1200, U.S.A.}
\author{George Siopsis}
\email{siopsis@tennessee.edu}
\affiliation{Department of Physics and Astronomy, The University of Tennessee, Knoxville, Tennessee 37996-1200, U.S.A.}
\author{Christian Weedbrook}
\affiliation{CipherQ, 10 Dundas St E, Toronto, M5B 2G9, Canada}

\date{\today}

\begin{abstract}
Machine learning is a fascinating and exciting field within computer science. Recently, this excitement has been transferred to the quantum information realm. Currently, all proposals for the quantum version of machine learning utilize the finite-dimensional substrate of discrete variables. Here we generalize quantum machine learning to the more complex, but still remarkably practical, infinite-dimensional systems. We present the critical subroutines of quantum machine learning algorithms for an all-photonic continuous-variable quantum computer that achieve an exponential speedup compared to their equivalent classical counterparts. Finally, we also map out an experimental implementation which can be used as a blueprint for future photonic demonstrations.
\end{abstract}

\maketitle


\section{Introduction}

We are now in the age of big data~\cite{IBM}. An unprecedented era in history where the storing, managing and manipulation of information is no longer effective using previously used computational tools and techniques. To compensate for this, one important approach in manipulating such large data sets and extracting worthwhile inferences, is by utilizing machine learning techniques. Machine learning~\cite{Hastie2013,Mackay2003,Alpaydin2004,Bishop2007,Murphy2007} involves using specially tailored `learning algorithms' to make important predictions in fields as varied as finance, business, fraud detection, and counter terrorism. Tasks in machine learning can involve either supervised or unsupervised learning and can solve such problems as pattern and speech recognition, classification, and clustering. Interestingly enough, the overwhelming rush of big data in the last decade has also been responsible for the recent advances in the closely related field of artificial intelligence~\cite{LeCun2015}; with the achievements of AlphaGo being a remarkable milestone.

Another important field in information processing which has also seen a significant increase in interest in the last decade is that of quantum computing~\cite{Ladd2010}. Quantum computers are expected to be able to perform certain computations much faster than any classical computer. In fact, quantum algorithms have been developed which are exponentially faster than their classical counterparts~\cite{Shor1997,Lloyd23081996,PhysRevLett.83.5162,PhysRevE.59.2429,Dowling2006,Buluta02102009,You2011,PhysRevA.85.062304,PhysRevA.85.030304,PhysRevLett.103.150502}. Recently, a new subfield within quantum information has emerged combining ideas from quantum computing with artificial intelligence to form quantum machine learning. Such discrete-variable schemes have shown exponential speedup in learning algorithms, such as supervised and unsupervised learning~\cite{Lloyd2013}, support vector machine~\cite{Rebentrost2014}, cluster assignment~\cite{Lloyd2014} and others~\cite{Harrow2009,Aimeur2013,Pudenz2013,Hentschel2010,Wiebe2015}. Initial proof-of-principle experimental demonstrations have also been performed~\cite{Cai2013,Barz2014,Cai2015,Li2015}.

In this paper, we have developed learning algorithms based on a different, but equally important, type of substrate in quantum computing, those of continuous variables (CVs)~\cite{Braunstein2005,Weedbrook2011}. A CV system is characterized by having an infinite-dimensional Hilbert space described by measuring variables with a continuous eigenspectra. The year 1999 saw the first important attempt at developing a CV model of quantum computing~\cite{Lloyd1999}. Seven years later, the cluster state version~\cite{Raussendorf2001} of CVs~\cite{Zhang2006,Menicucci2006}, accelerated the field's interest due to experimental interest. The result of this were proof-of-principle demonstrations~\cite{Yokoyama2014,Miyata2014,Pysher2011,Takeda2013}, which culminated in an `on-the-run' one-million-node cluster~\cite{Yoshikawa2016,Yokoyama2013}, as well as a 60-node `simultaneous' cluster~\cite{Chen2014}. Further important theoretical work was also carried out~\cite{Marshall2014,Lau2013,Loock2007,Gu2009,Alexander2014,Demarie2014,Menicucci2015,Wang2014,Menicucci2011,Marshall2015v2}, including an important CV architecture that was finally fault tolerant~\cite{Menicucci2014}.

Here, we take advantage of the practical benefits of CVs (high-efficiency room-temperature detectors, broad bandwidths, large-scale entanglement generation, etc.) by generalizing quantum machine learning to the infinite dimension. Specifically, we develop the important CV tools and subroutines that form the basis of the exponential speedup in various machine learning algorithms. This includes matrix inversion, principle component analysis and vector distance. Furthermore, each of these crucial subroutines are given a finite squeezing analysis for future experimental demonstrations along with a suggested photonic implementation.


\section{Quantum Machine Learning for Continuous Variables}

\textit{Encoding State---}
The general quantum state of a $n$-mode system is given by
\be
|f\rangle = \int f(q_1, \ldots, q_n) |q_1\rangle\otimes \ldots |q_n\rangle dq_1\ldots dq_n~. \nonumber
\ee
If we use this state to encode a discrete set of classical data, $\mathbf{a}\equiv\{ a_x; x=1,\dots, N \}$, which requires at least $N$ classical memory cells, only $n=\log_d N$ modes are sufficient, i.e.,
\be\label{eq1Psi}
f_\mathbf{a}(q_1, \ldots, q_n)=\sum_{x=1}^N a_x \prod_{i=1}^n \psi_{x_i}(q_i)
\ee
where $d$ is the number of basis state in each mode;
$x=(x_1 x_2 \ldots x_n)$ is a $d$-nary representation of $x$; $\psi_j(q) \equiv \langle q | \psi_j\rangle$ for $j=1,\dots,d$ is the wavefunction of the $j$th single mode basis state, $|\psi_j\rangle$.  Here we assume the vector $\mathbf{a}$ is normalised.

Obtaining the classical value of each data $a_x$ still requires $\mathcal{O}(N)$ copies of $|f_\mathbf{a}\rangle$.  Nevertheless in some applications only the global behavior of the data set is interesting.  For example, the value $\langle f_\mathbf{a} | \hat{F}|f_\mathbf{a}\rangle$ can be computed efficiently by a quantum computer with significantly fewer copies of  $|f_\mathbf{a}\rangle$ \cite{Aaronson2015}.  Quantum machine learning algorithms take advantage of this property to reduce the amount of memory and operations needed.


If the data set $\mathbf{a}$ is sufficiently uniform, it is known that $|f_\mathbf{a}\rangle$ can be efficiently generated.  As an illustration, we outline in the Supplementary Section an explicit protocol to generate a state with $d=2$ coherent basis states, $|\psi_1\rangle=|\alpha\rangle$ and $|\psi_2\rangle = |-\alpha\rangle$.  Our protocol generalizes the discrete variable method in Ref. \cite{Soklakov2006} to CV system by utilizing the CV implementation of the Grover's operators, $e^{i \phi |\psi\rangle \langle \psi|}$ for any given $|\psi\rangle$, as well as the efficient generation of Cat states and coherent states \cite{Furusawa2011}.

The encoding state construction of general non-uniform data could be constructed by extending the discrete-variable quantum RAM (qRAM)~\cite{Giovannetti2008} to a CV system, or by using a hybrid scheme~\cite{Andersen2015a}, although the state generation efficiency of such a general encoded state remains an open question \cite{Note,Aaronson2015}.  Nevertheless, the versatility of CV machine learning is not limited to process classical data sets that involve a discrete number of data.  In the context of universal CV quantum computation, the output of a computer is a CV state that evolves under an engineered Hamiltonian \cite{Lloyd1999}; the wave function of such a full CV output cannot be expressed in the form of Eq. (\ref{eq1Psi}).  As we will see, the CV machine learning subroutines are capable of processing even full CV states, and they are thus more powerful than the discrete variable counterparts~\cite{Note2}.


\textit{Exponential swap gate ---}  In both the data state construction and the quantum machine learning operation, the generalized Grover's operator, $e^{i\rho' t}$, plays the main role of inducing a phase shift according to an ensemble of unknown given states $\rho'$.  As suggested in Ref.~\cite{Lloyd2014}, such an operation can be implemented by repeatedly applying the exponential swap operation and tracing out the auxiliary mode, i.e.,
\be \label{eq:erho}
\mathrm{tr}_{\rho'} (e^{i\delta t \mathcal{S} } \rho \otimes \rho' e^{-i\delta t \mathcal{S} })  = e^{i\delta t\rho'} \rho e^{-i\delta t \rho'} + \mathcal{O} (\delta^2)~,
\ee
where by definition the swap operator functions as  $\mathcal{S}|\psi_1\rangle|\psi_2\rangle = |\psi_2\rangle|\psi_1\rangle$.

Here we outline the procedure of implementing the exponential operator with standard CV techniques.  First of all, we need a qubit as control, which can be implemented by two auxiliary modes, 1 and 2, with one and only one photon in both modes, i.e., the state of the modes is $\cos\theta |01\rangle + i \sin \theta |10\rangle$.  The rotation angle $\theta$ is controllable by applying the rotation operator $R(\theta) \equiv e^{i \theta (\hat{a}_1\hat{a}_2^\dag + \hat{a}_1^\dag\hat{a}_2)}$, which can be implemented by linear optics~\cite{Furusawa2011}.  In addition, we need a controlled-swap operation,
\be\label{eqn3}
C^{c c'}_\mathcal{S}= e^{-\frac{\pi}{4} (\hat{a}_c \hat{a}_{c'}^\dag-\hat{a}_c^\dag \hat{a}_{c'})} e^{i \pi \hat{a}_1^\dag\hat{a}_1 \hat{a}^\dag_c \hat{a}_c}
e^{\frac{\pi}{4} (\hat{a}_c \hat{a}_{c'}^\dag-\hat{a}_c^\dag \hat{a}_{c'})}
\ee
which swaps the modes $c$ and $c'$ depending on the photon number of the control qubit.  The operations in $C^{c c'}_\mathcal{S}$ can be implemented with the quartic gate introduced in \cite{Marshall2014,Marshall2015v2}. See Appendix B for more detail.

The control qubit is first prepared in $|+\rangle\equiv(|01\rangle+|10\rangle)/\sqrt{2}$.  By applying the operations in sequence $\exp(i \theta \mathcal{S}) = C^{c c'}_\mathcal{S}R(\theta)C^{c c'}_\mathcal{S}$, the state becomes
\bea \label{eq:eswap}
&&C^{c c'}_\mathcal{S}R(\theta)C^{c c'}_\mathcal{S}|+\rangle|\psi\rangle_c |\phi\rangle_{c'} = |+\rangle e^{i\theta \mathcal{S}_{cc'}} |\psi\rangle_c |\phi\rangle_{c'}\nonumber\\
&\equiv& |+\rangle(\cos \theta |\psi\rangle_c |\phi\rangle_{c'} + i \sin \theta |\phi\rangle_c |\psi\rangle_{c'})~.
\eea
The method can be generalized to implement a multi-mode exponential swap, $\exp(i \theta \mathcal{S}_{cc'}\mathcal{S}_{dd'}\ldots)$, by applying $C^{c c'}_\mathcal{S}C^{d d'}_\mathcal{S}\ldots R(\theta)C^{c c'}_\mathcal{S}C^{d d'}_\mathcal{S}\ldots$.  We note that the precious resources of a single photon state is not measured or discarded, so it can be reused in future operations.

We emphasize that, in stark contrast to the proposed implementation of exponential-swap gate in \cite{Lloyd2014} which is \textit{logical} and thus composed by a series of discrete variable logic gates, our implementation of the exponential-swap gate is \textit{physical}, i.e., it can be applied to full CV states that could not be written as the discrete variable form in Eq. (\ref{eq1Psi}).  This property allows our subroutine to be applied in, e.g. quantum tomography of CV states, which is more complicated than the discrete variable counterparts due to the large degree of freedom.

\section{CV Quantum Machine Learning Algorithms}

Now we discuss several key subroutines (matrix inversion, principle component analysis, and vector distance) that power the quantum machine learning problems using the tools we have just introduced.

\textit{Matrix inversion ---}  Various machine learning applications involves high-dimensional linear equations, e.g., $\mathbf{A}\mathbf{y}=\mathbf{b}$.  The advantage of some quantum machine learning algorithms is the ability to solve linear equations efficiently.  Specifically, for any vector $\mathbf{b} = \sum_i b_i \mathbf{e}_i$, computing the solution vector $\mathbf{y}=\mathbf{A}^{-1}\mathbf{b} = \sum_i b_i/\lambda_i \mathbf{e}_i$ is more efficient on a quantum computer~\cite{Harrow2009}.

In a CV system, the algorithm starts by preparing the state $|\mathbf{b}\rangle$ and two auxiliary modes in the $q$ quadrature eigenstates, i.e., $|0\rangle_{q,\mathcal{R}}$ and $|0\rangle_{q,\mathcal{S}}$.  We apply the operator $\exp(i \delta \gamma \mathbf{A} \hat{p}_\mathcal{R} \hat{p}_\mathcal{S} )$ $1/\delta$ times.  Each operator can be implemented based on Eq.~(\ref{eq:erho}), and a modified exponential swap gate with the rotation operator in Eq.~(\ref{eq:eswap}) replaced by the four-mode operator
\be
R(\gamma \hat{p}_\mathcal{R} \hat{p}_\mathcal{S}) =  e^{i\gamma \hat{p}_\mathcal{R} \hat{p}_\mathcal{S} (\hat{a}_1\hat{a}_2^\dag + \hat{a}_1^\dag\hat{a}_2)}~,
\ee
which can be implemented efficiently \cite{Marshall2014}. The state then becomes
\be
e^{i \gamma \mathbf{A} \hat{p}_\mathcal{R} \hat{p}_{\mathcal{S}} } |\mathbf{b}\rangle |0\rangle_{q,\mathcal{R}} |0\rangle_{q,\mathcal{S}} =\sum_i b_i \int |\mathbf{e}_i\rangle |p\rangle_{p,\mathcal{R}} |\gamma \lambda_i p \rangle_{q,\mathcal{S}} dp,
\ee
where we have neglected a normalization constant.  If the $\mathcal{S}$ auxiliary mode is measured in the $q$ quadrature with outcome $q_\mathcal{S}$, then we get
\be
\sum_i b_i/\lambda_i |\mathbf{e}_i\rangle |q_\mathcal{S}/\gamma\lambda_i\rangle_{p,\mathcal{R}}~.
\ee
Up to the normalization, the solution state $|\mathbf{y}\rangle = \sum_i b_i/\lambda_i |\mathbf{e}_i\rangle$ is obtained if the $\mathcal{R}$ auxiliary mode is measured in the $q$ quadrature and we get the result $q_\mathcal{R}=0$.

In the infinitely squeezed case, the successful rate of the last measurement is vanishing.  In practice, however, when squeezed vacuum states are employed as auxiliary modes, the successful rate of obtaining an answer state with error $\epsilon$ scales as $\mathcal{O}(\epsilon^{3/2})$, which is comparable to the discrete-variable algorithm that has success which scales as $\mathcal{O}(\epsilon)$~\cite{Rebentrost2014}.  The detailed argument is shown in Appendix C.

\textit{Principal component analysis ---} The next problem is to find the eigenvalue $\lambda$ corresponding to a unit eigenvector $\mathbf{e}_i$ with respect to the matrix $\mathbf{A}$, i.e., $\mathbf{A}\mathbf{e}_i=\lambda_i \mathbf{e}_i$.  This problem is ubiquitous in science and engineering and can also be used in quantum tomography, supervised learning and cluster assignment.

The algorithm starts from a data state $|\mathbf{e}_i\rangle$ and an auxiliary mode $\mathcal{R}$ prepared as the zero eigenstate of the $q$ quadrature, $|0\rangle_{q,\mathcal{R}}$.  The idea of the algorithm is to apply the operator $e^{i\gamma \mathbf{A}\hat{p}_R}$ that displaces the auxiliary mode according to the eigenvalue, i.e.,
\be e^{i\gamma \mathbf{A}\hat{p}_\mathcal{R}} |\mathbf{e}_i\rangle |0\rangle_{q,\mathcal{R}}=  |\mathbf{e}_i\rangle e^{i\gamma \lambda_i \hat{p}_\mathcal{R}} |0\rangle_{q,\mathcal{R}} = |\mathbf{e}_i\rangle |\gamma \lambda_i\rangle_{q,\mathcal{R}}~, \ee
then the eigenvalue can be obtained by measuring the auxiliary mode with homodyne detection.  This operator can be implemented by preparing an ensemble such that the density matrix is $\rho' = \mathbf{A}/\textrm{tr}\mathbf{A}$, and repeatedly apply the techniques in Eq. (\ref{eq:erho}) to implement $e^{i \delta \mathbf{A}\hat{p}_\mathcal{R}}$, for $\gamma \textrm{tr}\mathbf{A} /\delta$ times.  Here the argument of the exponential swap operator is not a c-number but an operator $\hat{p}_\mathcal{R}$.  This can be implemented by replacing the rotation operator in Eq. (\ref{eq:eswap}) by the three-mode operator
\be
R(\hat{p}_R) =  e^{i \delta \hat{p}_\mathcal{R} (\hat{a}_1\hat{a}_2^\dag + \hat{a}_1^\dag\hat{a}_2)}~,
\ee
which can be efficiently implemented by a cubic phase gate and linear optics \cite{Marshall2014, Lloyd1999}.

In practice, the success of the algorithm relies on the distinguishability of $|\gamma \lambda_i\rangle_q$, which depends on the spectrum of eigenvalues, the degree of squeezing $s$ of the auxiliary state, and the magnitude of error.  In Appendix~D, we have shown that $\mathcal{O}(1/\epsilon)$ operations are needed for an error $\epsilon \lesssim 1/(\gamma^2s)$.

\textit{Vector distance ---}  In supervised machine learning, new data is categorized into groups by its similarity to the previous data.  For example, the belonging category of a vector $\mathbf{u}$ is determined by the distance, $D$, to the average value of the previous data $\{\mathbf{v}_i \}$.  The objective of a quantum machine learning algorithm is to compute the value $D^2 \equiv |\mathbf{u} - \sum_{i=1}^M \mathbf{v}_i/M|^2$.

Following the approach given in Ref.~\cite{Rebentrost2014}, we assume an oracle can generate the state %
\be
|\Psi\rangle = \frac{1}{\mathcal{N}}\Big(|\mathbf{u}||0\rangle_I |\mathbf{\tilde{u}}\rangle + \frac{1}{\sqrt{M}}\sum_{i=i}^M |\mathbf{v}_i||i\rangle_I |\mathbf{\tilde{v}}_i\rangle\Big)~,
\ee
where the first mode is denoted as the index mode $I$; the normalization $\mathcal{N} \equiv \sqrt{|\mathbf{u}|^2+\sum_i |\mathbf{v}_i|^2/M}$ is supposed to be known.  $D^2$ can be obtained by conducting a swap test on the index mode with a reference mode prepared as $|\Phi\rangle_\mathcal{R} \equiv (|0\rangle_\mathcal{R} -\sum_{i=1}^M |i\rangle_\mathcal{R}/\sqrt{M})/\sqrt{2}$.  Various swap tests for CV systems have been proposed where the result is obtained from a photon number measurement~\cite{Filip2002, Jeong2014}.  Here we propose a swap test that employs only homodyne detection and an exponential swap operation.

We consider two test modes that are prepared in the coherent states $|\beta\rangle_1|0\rangle_2$.  The operator $\exp(i\frac{\pi}{4}\mathcal{S}_{12}\mathcal{S}_{I\mathcal{R}})$ is applied to exponential swap the two test modes, as well as the reference and the index modes.  After that, the test modes pass through a $50/50$ beam splitter.  The density operator of the test modes after tracing out the other modes becomes
\bea
\rho_{12} &= & \frac{1}{2}\Big(|\frac{\beta}{\sqrt{2}} \rangle_{11} \langle\frac{\beta}{\sqrt{2}} | + i D^2 |\frac{\beta}{\sqrt{2}}\rangle_{11} \langle \frac{-\beta}{\sqrt{2}} |   \\
&&- i D^2 |\frac{-\beta}{\sqrt{2}} \rangle_{11} \langle \frac{\beta}{\sqrt{2}}| +|\frac{-\beta}{\sqrt{2}}\rangle_{11} \langle\frac{-\beta}{\sqrt{2}} |\Big)\otimes |\frac{\beta}{\sqrt{2}}\rangle_{22}\langle\frac{\beta}{\sqrt{2}}|~.\nonumber
\eea
We find that if the $1$ mode is homodyne detected in the $p$ quadrature and $\beta \gtrsim 4$, the probability difference of measuring a positive and negative outcome scales as $D^2$, where the scaling constant is at the order of $0.1$ for a wide range of $\beta$. See Appendix E for further details.

\begin{figure*}
\includegraphics[width=5in]{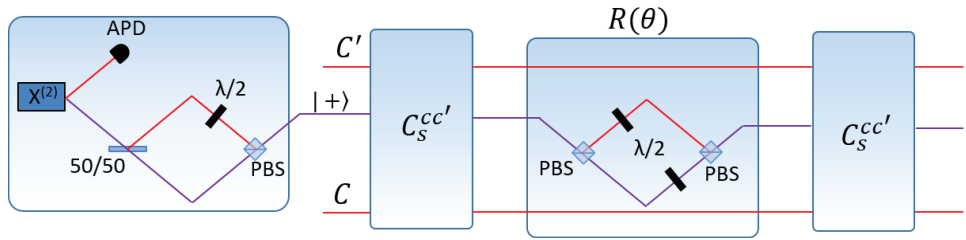}
\caption{All-photonic implementation schematic of the operator $\exp(i \theta \mathcal{S}) = C^{c c'}_\mathcal{S}R(\theta)C^{c c'}_\mathcal{S}$. We initially have an ancillary input mode $|+\rangle =(|01\rangle+|10\rangle)/\sqrt{2}$ with two (swap) modes $C$ and $C'$ used to implement the operators given in Eq.~\ref{eq:eswap}. The method for generating $|+\rangle$ is one of many possibilities, e.g., preparing a heralded superposition of polarization states is illustrated. $\chi^{(2)}$: nonlinear crystal source; APD: avalanche photodiode detector; $50/50$: balanced beam splitter; $\lambda/2$: half wave plate; PBS: polarizing beam splitter; $C_S^{CC'}$: controlled-swap operator; $R (\theta)$: rotation operator; see text for explanation of operators. Note that $C^{c c'}_\mathcal{S}$ can be implemented with the quartic gate~\cite{Marshall2014,Marshall2015v2} and $R(\theta)$ can also be efficiently implemented using linear optics.}\label{fig1.eps}
\end{figure*}

\section{All-Photonic Implementation}

In this section, we outline an all-photonic implementation of the previously mentioned machine learning algorithms. First, one must create an ancillary state for use in the exponential swap gate. One method is to provide a heralded ancilla via parametric down conversion~(see for example \cite{Furusawa2011}, for background on a lot of the standard quantum optics methods discussed here). The undetected photon is interfered with the vacuum on a $50/50$ beam splitter in order to place it in the superposition required for Eq.~(\ref{eq:eswap}) (see Fig.~\ref{fig1.eps}). This serves as an input to the phase-dependent gates outlined in~\cite{Marshall2014}, which can be used to construct the exponential swap gate. The rotation gate in Eq.~(\ref{eq:eswap}) is essentially the interference of the two modes on a variable reflectivity, or programmable beam splitter, which can be achieved via polarization control and a polarizing beam splitter, or via a collection of phase or amplitude modulators. Inverse phase-dependent gate operations are implemented after the rotation.

Each algorithm essentially utilizes a variation of this configuration, in addition to the possibility of squeezed ancilla in order to increase the accuracy of the result. The principle component analysis problem replaces the variable beam splitter in the swap gate with a two-mode quantum-non-demolition phase gate. It can be implemented by treating the auxiliary mode, $\mathcal{R}$, as the ancilla in the phase-dependent gate. Thus, the principle component analysis problem essentially relies on repeated application of the `repeat-until-success' phase gate~\cite{Marshall2014}. In a realistic scenario, $\mathcal{R}$ is in a single-mode squeezed state with finite squeezing (see Appendix C.2), which is experimentally straightforward using a below-threshold optical parametric amplifier (OPA). Phase sensitive amplification can also be used. The squeezing parameter can be used to tune the accuracy of the computation. The final homodyne detection is also experimentally straight forward with a local oscillator derived from the pump laser used in the OPA (via a doubling cavity, for instance).

The matrix inversion algorithm is experimentally very similar to the eigenvalue problem. The key difference is the use of an extra auxiliary mode, which can be prepared independently with an additional OPA. The four-mode operator is conceptually similar to the operator in Eq.~(6) used in the previous algorithm. Each auxiliary mode serves as an ancilla in the phase-dependent gate, and the algorithm otherwise follows a similar approach to the previous one, with a final homodyne detection step for the amplitude quadrature of each auxiliary mode, with the local oscillators derived from the pumps of each OPA.

Finally, the vector distance algorithm requires use of a swap test, which can be implemented via the application of the exponential swap gate between two auxiliary states (which can be coherent states or squeezed states) and the oracle mode in Eq.~(10)~\cite{Rebentrost2014} and the reference mode. The required homodyne detection of the phase quadrature of the first test mode in a bright coherent state and is again experimentally straight forward.

\section{Discussion}

The previous all-photonic implementations are difficult to do experimentally but are still within current reach of the latest technological achievements. For instance, high rates of squeezing are now achievable~\cite{Andersen2015}, along with the generation of cat states~\cite{Gao2010}. However, we note that our scheme is not limited to photonic demonstrations but a variety of substrates, including spin ensemble systems, such as trapped atoms and solid state defect centers~\cite{Dowling1994,Tordrup2008,Wesenberg2009,Kubo2010,Kubo2011}.

We hope that the work presented here will lead to further avenues of research. Especially since there has been a substantial increase of results in discrete-variable machine learning~\cite{Wiebe2012,Wiebe2014,Paparo2014,Wiebe2015,Amin2016,Wiebe2016}. All of these would be interesting to be generalized to continuous variables as future work. Additionally, adapting our current work into the cluster-state formulism~\cite{Gu2009} would also be interesting in order to take advantage of state-of-the-art experimental interest and the scalability that continuous variables can provide~\cite{Yokoyama2013}. Furthermore, we note another viable option that uses a `best-of-both-worlds' approach to quantum information processing, i.e., hybrid schemes~\cite{Liu2015,Andersen2015a,LauPlenio2016}. It would be interesting to adapt our scheme presented here to such hybrid architectures.
%

\acknowledgments We thank Kevin Marshall for helpful discussions. H.-K.~L would like to acknowledge support from the Croucher Foundation. R.~C.~P. performed portions of this work at Oak Ridge National Laboratory, operated by UT-Battelle for the US Department of Energy under Contract No.\ DE-AC05-00OR22725.

\appendix

\section{Appendix}

\section{Encoding efficiency \label{sec:encoding}}

Here we discuss the encoding efficiency as described in Sec.~II. Of particular interest are the Grover operators,
\be e^{i\pi |\Psi_0\rangle\langle \Psi_0|} = \mathbb{I} - 2 |\Psi_0\rangle \langle \Psi_0| \ee
and $e^{i\phi |x\rangle\langle x|}$ (which implements a phase change $e^{i\phi}$ on the state $|x\rangle$). A repeated application of these unitaries can create any state $|\Psi\rangle$ from $|\Psi_0\rangle$. The complexity $\mathcal{C}$ (number of resources and oracle calls required) varies depending on the distribution of data. For probability distributions which remain uniformly bounded for large $N$, the complexity has a polynomial dependence on $n$. More precisely, this is the case if $|a_x| \lesssim 1/\sqrt{N}$, $\forall x$ (for example, states close to $|\Psi_0\rangle$). Otherwise, the complexity can be as high as $\mathcal{O} (\sqrt{N})$. The latter is the case for any state, such as $|x\rangle$, in which the probability distribution is highly peaked (although, it should be noted that a state $|x\rangle$ is easy to construct, because it is the tensor product of coherent states). The proof follows the lines of Ref.~\cite{Soklakov2006}. The number $1/\epsilon$ of copies of $|\Psi_0\rangle$ and $|x\rangle$ needed for implementation of $e^{i\pi |\Psi_0\rangle\langle \Psi_0|}$ and $e^{i\phi |x\rangle\langle x|}$ are also polynomial in $n$. Indeed, if $\lambda$ is the required fidelity, we have $\lambda \sim \mathcal{C}\epsilon$, therefore the number of copies of $|\Psi_0\rangle$ and $|x\rangle$ needed is $1/\epsilon \sim \mathcal{C}/\lambda$, which is of polynomial order in $n$.

\section{Higher-order CV non-Gaussian gates}\label{app2nongauss}

Here we discuss the implementation of higher-order gates using CVs. The discussion generalizes the construction of cubic phase gate given in Ref.~\cite{Marshall2014}. Non-Gaussian phase gates of order $k$ are of the form $e^{i\gamma P_k(\hat{x})}$, where $P_k(\hat{x})$ is a polynomial of order $k$ ($k>2$). In this paper, we make use of cubic ($k=3$) and quartic ($k=4$) phase gates. To implement them, we first decompose them as
\be e^{i\gamma P_k(\hat{x})} = \left( 1 + i\frac{\gamma}{K} P_k(\hat{x}) \right)^K + \mathcal{O} (1/K)\ee
and further,
\be 1 + i\frac{\gamma}{K} P_k(\hat{x}) = \mathcal{U}_0 \mathcal{U}_1 \cdots \mathcal{U}_{k-1} \ee
where $\mathcal{U}_l = 1 + \gamma_l \hat{x}$, and $-1/\gamma_l$ are the (complex) roots of the $k$th-order polynomial $1 + i\frac{\gamma}{K} P_k(\hat{x})$. Each linear operator $\mathcal{U}_l$ ($l=0,1,\dots,k-1$) can be implemented as discussed in further detail in Ref.~\cite{Marshall2014}.

Specifically, the quartic gate needed for the controlled-swap operator $C_{\mathcal{S}}^{cc'}$ can be written as
\be U_{\mathcal{S}} = e^{i\pi H_1H_c } \ee
where $H_1 = \hat{p}_1^2 + \hat{x}_1^2$ and
$H_c = \hat{p}_c^2 + \hat{x}_c^2$.
To implement it, we decompose it as
\be U_{\mathcal{S}} = \left( e^{\frac{i\pi}{K} \hat{p}_1^2 \hat{p}_c^2 } e^{\frac{i\pi}{K} \hat{p}_1^2 \hat{x}_c^2 } e^{\frac{i\pi}{K} \hat{x}_1^2 \hat{p}_c^2 } e^{\frac{i\pi}{K} \hat{x}_1^2 \hat{x}_c^2 } \right)^K + \mathcal{O} (1/K) \ee
The first three factors can written in terms of the last factor, respectively, as
\bea
e^{\frac{i\pi}{K} \hat{p}_1^2 \hat{p}_c^2 } &=& e^{\frac{i\pi}{2} H_1 }e^{\frac{i\pi}{2} H_c} e^{\frac{i\pi}{K} \hat{x}_1^2 \hat{x}_c^2 } e^{-\frac{i\pi}{2} H_c}e^{-\frac{i\pi}{2} H_1}\nonumber\\
e^{\frac{i\pi}{K} \hat{p}_1^2 \hat{x}_c^2 } &=& e^{\frac{i\pi}{2} H_1 } e^{\frac{i\pi}{K} \hat{x}_1^2 \hat{x}_c^2 } e^{-\frac{i\pi}{2} H_1}\nonumber\\
e^{\frac{i\pi}{K} \hat{x}_1^2 \hat{p}_c^2 } &=& e^{\frac{i\pi}{2} H_c } e^{\frac{i\pi}{K} \hat{x}_1^2 \hat{x}_c^2 } e^{-\frac{i\pi}{2} H_c}
\eea
The last factor can be written in terms of quartic gates as
\bea\label{eqlast} e^{\frac{i\pi}{K} \hat{x}_1^2 \hat{x}_c^2 } &=& e^{2ip_1 x_c}e^{\frac{i\pi}{12K} \hat{x}_1^4 }e^{-4ip_1x_c}e^{\frac{i\pi}{12K} \hat{x}_1^4 } e^{2ip_1x_c}e^{-\frac{i\pi}{6K} \hat{x}_1^4 } \nonumber\\
&&\times e^{-\frac{i\pi}{6K} \hat{x}_c^4 }
\eea
The other non-Gaussian gates used in our calculations can be implemented similarly. The strategy is to first decompose the operator into simple factors with an error $\mathcal{O} (1/K)$, and then rotate each $\hat{p}_i$ into $\hat{x}_i$ using $e^{\frac{i\pi}{2} (\hat{p}_i^2 + \hat{x}_i^2)}$. Thus, we obtain an exponent which is a polynomial in $\hat{x}_i$. The latter can be written in terms of single-mode unitaries, as in \eqref{eqlast}.

It should be noted that it is not necessary to introduce an error $\mathcal{O} (1/K)$, because the operators considered here have exponents which are polynomial in $\hat{x}_i$ and $\hat{p}_i$. Therefore, it is possible to perform an exact decomposition of these operators into simple factors. We will not do this here, because the expressions become long and do not serve our purpose of demonstrating the implementation of these unitaries using cubic and quartic gates~\cite{Marshall2014}.

\section{Matrix inversion}

\subsection{Ideal implementation}

We start with qumodes in the state $|b\rangle$ and two resource modes in the state $|q_\mathcal{R} = 0\rangle\otimes|\tilde{q}_\mathcal{R} = 0\rangle$. Thus, initially,
\be |\Psi\rangle \propto |b\rangle \int dp_\mathcal{R} d\tilde{p}_\mathcal{R} |p_\mathcal{R}\rangle |\tilde{p}_\mathcal{R}\rangle \ee
%
Next, we apply the unitary
\be\label{eq40} e^{i\gamma Ap_\mathcal{R} \tilde{p}_\mathcal{R}} \ee
where $\gamma$ is a parameter that can be adjusted at will. The unitary is implemented
similarly to \eqref{eq19}, except that we now have two resource qumodes. The algorithm is unchanged, except for the rotation of the ancillas, which becomes a four-mode unitary,
$R_{AB} \left[ \gamma \frac{p_\mathcal{R} \tilde{p}_\mathcal{R}}{n} \right]$,
and can be implemented via the quartic gate constructed in~\cite{Marshall2014}.

We obtain
\be e^{i\gamma Ap_\mathcal{R} \tilde{p}_\mathcal{R}} |\Psi\rangle \propto \sum_i \beta_i \int dp_\mathcal{R} d\tilde{p}_\mathcal{R} e^{i\gamma \lambda_i p_\mathcal{R} \tilde{p}_\mathcal{R}}   |e_i\rangle|p_\mathcal{R}\rangle |\tilde{p}_\mathcal{R}\rangle\ee
Next, we measure $q_\mathcal{R}$ of the first resource mode. This projects the state onto
\bea |q_\mathcal{R}\rangle\langle q_\mathcal{R}|e^{i\gamma Ap_\mathcal{R} \tilde{p}_\mathcal{R}} |\Psi\rangle &\propto& \sum_i \beta_i \int d\tilde{p}_\mathcal{R}   \delta(\gamma\lambda_i \tilde{p}_\mathcal{R} - q_\mathcal{R}) \nonumber\\
&&\times |e_i\rangle|q_\mathcal{R}\rangle |\tilde{p}_\mathcal{R}\rangle \nonumber\\
&\propto& \sum_i \frac{1}{\lambda_i}\beta_i  |e_i\rangle|q_\mathcal{R}\rangle \left| \tilde{p}_\mathcal{R} = \frac{q_\mathcal{R}}{\gamma\lambda_i} \right\rangle \nonumber\\
\eea
Finally, we measure $\tilde{q}_\mathcal{R}$. The final state is
\be\label{eq44} \sum_i e^{-i\tilde{q}_\mathcal{R} q_\mathcal{R}/(\gamma \lambda_i)}\frac{1}{\lambda_i}\beta_i  |e_i\rangle|q_\mathcal{R}\rangle \left| \tilde{q}_\mathcal{R} \right\rangle \ee
For sufficiently large $\gamma$, this is approximately
\be \sum_i \frac{1}{\lambda_i} \beta_i |e_i\rangle|q_\mathcal{R}\rangle |0\rangle  =
| A^{-1} b\rangle |q_\mathcal{R}\rangle |\tilde{q}_\mathcal{R}\rangle \ee

\subsection{Realistic implementation}

Realistically, we start with qumodes that are squeezed states. Thus, initially,
\be |\Psi\rangle \propto |b\rangle \int dp_\mathcal{R} d\tilde{p}_\mathcal{R} e^{-[(p_\mathcal{R})^2+ (\tilde{p}_\mathcal{R})^2]/(2s)}|p_\mathcal{R}\rangle |\tilde{p}_\mathcal{R}\rangle \ee
Ideally, $s \to \infty$. After we apply the unitary \eqref{eq40},
we obtain
\bea e^{i\gamma Ap^R \tilde{p}^R} |\Psi\rangle &\propto& \sum_i \beta_i \int dp^R d\tilde{p}^R e^{-[(p^R)^2+ (\tilde{p}^R)^2]/(2s)} \nonumber\\
&&\times e^{i\gamma \lambda_i p^R \tilde{p}^R}  |e_i\rangle|p^R\rangle |\tilde{p}^R\rangle\eea
After we measure $q_\mathcal{R}, \tilde{q}_\mathcal{R}$ of the respective resource modes, we arrive at the final state
\be e^{i\gamma Ap_\mathcal{R} \tilde{p}_\mathcal{R}} |\Psi\rangle \propto \sum_i \beta_i \mathcal{A}_i (q_\mathcal{R},\tilde{q}_\mathcal{R}) |e_i\rangle|q_\mathcal{R}\rangle |\tilde{q}_\mathcal{R}\rangle\ee
where
\bea \mathcal{A}_i (q_\mathcal{R},\tilde{q}_\mathcal{R}) &=& \int dp_\mathcal{R} d\tilde{p}_\mathcal{R} e^{-[(p_\mathcal{R})^2+ (\tilde{p}_\mathcal{R})^2]/(2s)} \nonumber\\
&&\times e^{i(\gamma \lambda_i p_\mathcal{R} \tilde{p}_\mathcal{R}-p_\mathcal{R} q_\mathcal{R}-\tilde{p}_\mathcal{R}\tilde{q}_\mathcal{R})} \nonumber\\
&\propto& \frac{\exp \left[ -\frac{s[(q_\mathcal{R})^2 + (\tilde{q})^2]+2is^2 \gamma\lambda_i q_\mathcal{R} \tilde{q}_\mathcal{R}}{2(1+\gamma^2 \lambda_i^2 s^2)}\right]}{\lambda_i\sqrt{1+ \frac{1}{\gamma^2\lambda_i^2 s^2}}} \eea
Notice that in the limit $s\to\infty$, this reduces to $\mathcal{A}_i (q_\mathcal{R},\tilde{q}_\mathcal{R}) \propto \frac{1}{\lambda_i}\exp\left[ \frac{iq_\mathcal{R} \tilde{q}_\mathcal{R}}{\gamma\lambda_i} \right]$, in agreement with \eqref{eq44}.

For $\gamma |\lambda_i| s \sim \frac{1}{\sqrt{\epsilon}}$, we have
\be |\mathcal{A}_i (q_\mathcal{R},\tilde{q}_\mathcal{R})|^2 \sim \frac{\exp \left[ -\frac{(q_\mathcal{R})^2 + (\tilde{q})^2}{2\gamma^2 \lambda_i^2 s}\right]}{\lambda_i} \ee
so both $q_\mathcal{R}$ and $\tilde{q}_\mathcal{R}$ have probability distributions of width $\sim \gamma |\lambda_i| \sqrt{s} \sim \mathcal{O} (1/\sqrt{s\epsilon})$.
The width of $\frac{q_\mathcal{R}\tilde{q}_\mathcal{R}}{\gamma\lambda_i}$ is $\sim \gamma |\lambda_i| s \sim\mathcal{O} (1/\sqrt{\epsilon})$. If we want $\frac{q_\mathcal{R}\tilde{q}_\mathcal{R}}{\gamma\lambda_i} \lesssim \mathcal{O} (\epsilon)$, for a normal distribution the success rate is $\mathcal{O} (\epsilon^{3/2})$.

\section{Eigenvalue distinguishing}

Given an $N\times N$ Hermitian matrix $A$, and a vector $|b\rangle$, find out if $|b\rangle$ is an eigenvector, and if so, which eigenvalue it belongs to. In particular, we are interested in matrices of the form $A = \rho - \sigma$, where $\rho,\sigma$ are both mixed states.

\subsection{Ideal implementation}

We start with $n$ qumodes in the state $|b\rangle$ and a resource mode $\mathcal{R}$ in the state $|q_\mathcal{R} =0\rangle$. Thus, initially,
\be |\Psi\rangle = |b\rangle \int dp_\mathcal{R} |p_\mathcal{R}\rangle \ee
Next, we apply the unitary
\be\label{eq19} e^{i\gamma Ap_\mathcal{R}} \ee
If the eigenvalue problem of $A$ is
\be A|e_i\rangle = \lambda_i |e_i\rangle \ee
and we expand
\be |b\rangle = \sum_i \beta_i |e_i\rangle \ee
we obtain
\be e^{i\gamma Ap_\mathcal{R}} |\Psi\rangle = \sum_i \beta_i \int dp_\mathcal{R} e^{i\gamma \lambda_i p_\mathcal{R}} |e_i\rangle |p_\mathcal{R}\rangle \ee
Next, we measure $q_\mathcal{R}$ of the resource mode. This projects the state onto
\bea \langle q_\mathcal{R}|e^{i\gamma Ap_\mathcal{R}} |\Psi\rangle &=& \sum_i \beta_i \int dp_\mathcal{R} e^{i(\gamma \lambda_i -q_\mathcal{R}) p_\mathcal{R}} |e_i\rangle \nonumber\\
&=& \sum_i \beta_i \delta (\gamma \lambda_i -q_\mathcal{R}) |e_i\rangle \eea
Thus, the measurement outcome is proportional to one of the eigenvalues, $q_\mathcal{R} = \lambda_i/\gamma$, for which $\beta_i \ne 0$. The most probable outcome corresponds to the maximum $|\beta_i|^2$. We obtain that outcome with certainty, if $|b\rangle$ is an eigenstate of $A$.

All of the above steps are independent of the size of the matrix, $N$. This is evident for all steps, except for the implementation of the unitary \eqref{eq19}. To implement \eqref{eq19}, we make $1/\epsilon$ copies of $\rho$ and $1/\epsilon$ copies of $\sigma$, where $\epsilon$ is the desired accuracy. Let $\mathcal{S}$ be the swap operator. We have
\bea \mathrm{tr}_P e^{i\epsilon \gamma \mathcal{S} p_\mathcal{R}} \rho \otimes |b\rangle\langle b| e^{-i\epsilon \gamma\mathcal{S} p_\mathcal{R}} &=& e^{i\epsilon \gamma \rho p_\mathcal{R}} |b\rangle\langle b| e^{-i\epsilon \gamma \rho p_\mathcal{R}} \nonumber\\
&& + \mathcal{O} (\epsilon^2) \eea
where we took a partial trace over the degrees of freedom of $\rho$. Similarly for $\sigma$,
\bea \mathrm{tr}_P e^{-i\epsilon \gamma \mathcal{S} p_\mathcal{R}} \sigma \otimes |b\rangle\langle b| e^{i\epsilon\gamma \mathcal{S} p_\mathcal{R}} &=& e^{-i\epsilon\gamma \sigma p_\mathcal{R}} |b\rangle\langle b| e^{i\epsilon\gamma\sigma p_\mathcal{R}} \nonumber\\
&& + \mathcal{O} (\epsilon^2) \eea
The smaller the $\epsilon$, the larger the range of $p_\mathcal{R}$ over which there is little distortion. Repeating these two steps with the rest of the copies, we arrive at an approximation of
\be e^{i\gamma A p_\mathcal{R}} |b\rangle\langle b| e^{-i\gamma A p_\mathcal{R}} \ee
i.e., an implementation of \eqref{eq19}.

The unitary $e^{i\epsilon\gamma\mathcal{S} p_\mathcal{R}}$ itself can be implemented using
\be\label{eq26} e^{i\epsilon\gamma\mathcal{S} p_\mathcal{R}} = \mathbb{I} \cos \epsilon \gamma p_\mathcal{R} + i\mathcal{S} \sin \epsilon \gamma p_\mathcal{R} \ee
To implement \eqref{eq26}, we introduce two ancillary modes in the logical state $|0\rangle_L$, and rotate it to $\frac{1}{\sqrt{2}} \left( |0\rangle_L + |1\rangle_L \right)$.
Next, we apply the string of three-mode unitaries $\mathbb{S}$,
and arrive at the state
\be \frac{1}{\sqrt{2}} \left( \mathbb{I} \otimes |0\rangle_1 |1\rangle_2 + \mathcal{S} \otimes |1\rangle_1 |0\rangle_2 \right) \ee
We then apply the three-mode unitary $R_{AB} (\epsilon \gamma p_\mathcal{R})$,
which is a rotation on the ancillas and can be implemented using the cubic gate constructed in \cite{Marshall2014} together with two-mode operators. The state becomes
\bea &&\frac{1}{\sqrt{2}} \left[ \mathbb{I} \otimes \left( \cos \epsilon \gamma p_\mathcal{R} |0\rangle_L +i\sin \epsilon \gamma p_\mathcal{R} |1\rangle_L \right) \right. \nonumber\\
&& \left. + \mathcal{S} \otimes \left( \cos \epsilon \gamma p_\mathcal{R} |1\rangle_L +i\sin \epsilon \gamma p_\mathcal{R} |0\rangle_L \right) \right] \eea
Once again, we apply $\mathbb{S}$ and obtain
\be \frac{1}{\sqrt{2}} \left[ \mathbb{I} \cos \epsilon \gamma p_\mathcal{R} +i\mathcal{S} \sin \epsilon \gamma p_\mathcal{R}   \right]
\otimes \left( |0\rangle_L + |1\rangle_L \right)\ee
After applying $R_{AB} (-\frac{\pi}{4})$, the ancilla goes back to its original state, and we arrive at
\be \left[ \mathbb{I} \cos \epsilon \gamma p_\mathcal{R} +i\mathcal{S} \sin \epsilon \gamma p_\mathcal{R}   \right]
\otimes |0\rangle_L \ee
matching \eqref{eq26}, as desired.

\subsection{Realistic implementation}

Realistically, we start with a qumode in the state $|b\rangle$ and a resource mode in a squeezed state. Thus, initially,
\be |\Psi\rangle \propto |b\rangle \int dp_\mathcal{R} e^{-(p_\mathcal{R})^2/(2s)}|p_\mathcal{R}\rangle \ee
After we apply the unitary \eqref{eq19},
and measure $q_\mathcal{R}$ of the resource mode, we obtain the projected state
\bea \langle q_\mathcal{R}|e^{i\gamma Ap_\mathcal{R}} |\Psi\rangle &\propto& \sum_i \beta_i \int dp_\mathcal{R} e^{-(p_\mathcal{R})^2/(2s)} e^{i(\gamma \lambda_i -q_\mathcal{R}) p_\mathcal{R}} |e_i\rangle \nonumber\\
&\propto& \sum_i \beta_i e^{- s(\gamma \lambda_i -q_\mathcal{R})^2/2} |e_i\rangle \eea
which yields a probability distribution
\be P(q_\mathcal{R}) \propto \sum_i |\beta_i|^2 e^{- s(\gamma \lambda_i -q_\mathcal{R})^2} \ee
consisting of peaks at the eigenvalues. Since we are interested in eigenvalues $\pm 1$, to discriminate between them, the width of the peaks ought to be $\mathcal{O}(1)$, so $s \gamma^2 \gtrsim 1$. The number $1/\epsilon$ of copies needed to simulate \eqref{eq19} must be such that $\epsilon s \gamma^2 \lesssim 1$, therefore by adjusting the arbitrary parameter $\gamma$, even a small integer $1/\epsilon$ will suffice.

\section{Distance computation}

Let $\mathbf{u}$ and $\mathbf{v}_i$ be two $N$-dimensional unit vectors.  We are interested in computing the distance $D$ between $\mathbf{u}$ to the average of $\{ \mathbf{v}_i \}$, i.e.,
\bea \label{eq1}
 D^2 \equiv |\mathbf{u} - \frac{1}{M}\sum_{i=1}^M\mathbf{v}_i|^2  &=& |\mathbf{u}|^2 + \frac{1}{M^2}\sum_{i,i'} |\mathbf{v}_i| |\mathbf{v}_{i'}| \mathbf{\tilde{v}}_i\cdot \mathbf{\tilde{v}}_{i'} \nonumber \\
&& - \frac{1}{M}\sum_{i=1}^M |\mathbf{u}||\mathbf{v}_i| (\mathbf{\tilde{u}}^\ast \cdot \mathbf{\mathbf{v}}_i +\mathbf{\tilde{u}} \cdot \mathbf{\mathbf{v}}_i^\ast) ~. \nonumber
\eea

The objective of quantum machine learning is to measure the value of $D^2$ without learning all of the coefficients of each data set.  Following~\cite{Rebentrost2014}, we consider a $n+1$ mode resources state, given by
\be
|\Psi\rangle = \frac{1}{\mathcal{N}}\Big(|\mathbf{u}||0\rangle_I |\mathbf{\tilde{u}}\rangle + \frac{1}{\sqrt{M}}\sum_{i=i}^M |\mathbf{v}_i||i\rangle_I |\mathbf{\tilde{v}}_i\rangle\Big)~,
\ee
where the normalization $\mathcal{N} \equiv \sqrt{|\mathbf{u}|^2+\sum_i |\mathbf{v}_i|^2/M}$ is supposed to be known.  We denote the first mode as the index mode $I$, while the following $n$ modes are the data modes.  Following our argument in Appendix \ref{sec:encoding}, if the data of $\mathbf{u}$ and $\mathbf{v}_i$ are sufficiently homogeneous, $|\Psi\rangle$ can also be efficiently constructed.

In analogous to the discrete-variable algorithm~\cite{Lloyd2013}, the value of $D^2$ can be deduced from the probability of measuring the index mode in the $|\Phi\rangle \equiv (|0\rangle -\sum_{i=1}^M |i\rangle/\sqrt{M})/\sqrt{2}$ state, i.e., $\|\langle\Psi|\Phi\rangle\|^2 = D^2/2\mathcal{N}^2$.  Such a measurement can be achieved by conducting a swap test between the index mode with an auxiliary reference mode that is prepared in $|\Phi\rangle_r$.  Here we propose a swap test that involves only homodyne detection and the exponential swap operation.

Let's consider two auxiliary test modes that are prepared in coherent states, $|\beta 0\rangle_{12}$.  Then an exponential swap $\exp(i\frac{\pi}{4}\mathcal{S}_{12}\mathcal{S}_{I_\mathcal{R}})$ is applied to transform the state as
\bea
&&\frac{1}{\sqrt{2}\mathcal{N}}\Big(|\beta 0\rangle_{12} (|\mathbf{u}| |\Phi\rangle_\mathcal{R} |0\rangle_I |\mathbf{\tilde{u}}\rangle + \frac{1}{\sqrt{M}}\sum_{i=i}^M |\mathbf{v}_i| |\Phi\rangle_\mathcal{R} |i\rangle_I |\mathbf{\tilde{v}}_i\rangle) \nonumber \\
&&+ i|0\beta\rangle_{12} (|\mathbf{u}||0\rangle_\mathcal{R} |\Phi\rangle_I |\mathbf{\tilde{u}}\rangle + \frac{1}{\sqrt{M}}\sum_{i=i}^M |\mathbf{v}_i| |i\rangle_\mathcal{R} |\Phi\rangle_I |\mathbf{\tilde{v}}_i\rangle) \Big)~. \nonumber
\eea
After tracing out the index mode, the reference mode, and the data modes, the total state of the two test modes becomes
\bea
\rho_{12} &=& \frac{1}{2}\Big(|\beta 0\rangle \langle \beta 0| - i \frac{D^2}{\mathcal{N}^2} |\beta 0\rangle \langle 0 \beta | \nonumber\\
&&+ i \frac{D^2}{\mathcal{N}^2} | 0 \beta\rangle \langle \beta 0 | + |0 \beta\rangle \langle 0 \beta| \Big)
\eea

After that, we apply a $50/50$ beam splitter that transforms any coherent state as $U_\textrm{BS}|\alpha\beta\rangle=|\frac{\alpha-\beta}{\sqrt{2}}\frac{\alpha+\beta}{\sqrt{2}}\rangle$, the state of the test mode then becomes
\bea
\rho_{12} &= & \frac{1}{2}\Big(|\frac{\beta}{\sqrt{2}} \rangle_{11} \langle\frac{\beta}{\sqrt{2}} | - i \frac{D^2}{\mathcal{N}^2} |\frac{\beta}{\sqrt{2}}\rangle_{11} \langle \frac{-\beta}{\sqrt{2}} |   \\
&&+ i  \frac{D^2}{\mathcal{N}^2} |\frac{-\beta}{\sqrt{2}} \rangle_{11} \langle \frac{\beta}{\sqrt{2}}| +|\frac{-\beta}{\sqrt{2}}\rangle_{11} \langle\frac{-\beta}{\sqrt{2}} |\Big)\otimes |\frac{\beta}{\sqrt{2}}\rangle_{22}\langle\frac{\beta}{\sqrt{2}}|~.\nonumber
\eea

When measuring the first test mode in the $P$ quadrature, the probability of obtaining a value $p$ is proportional to
\be \mathcal{P}(p) \propto e^{-p^2} (2 +i e^{i \sqrt{2} p \beta} \frac{D^2}{\mathcal{N}^2} - i e^{-i \sqrt{2} p \beta} \frac{D^2}{\mathcal{N}^2} )~.  \ee
We find that the probability difference between the positive and negative values of $p$ is
\be \mathcal{P}(p>0) - \mathcal{P} (p<0) = -e^{-\beta^2/2} \textrm{erfi}(\frac{\beta}{\sqrt{2}}) \frac{D^2}{\mathcal{N}^2}~,  \ee
where erfi is the imaginary error function.  For $\beta \gtrsim 4$, the expression is well approximated by $\sim 0.1 D^2/\mathcal{N}^2$.


\begin{thebibliography}{99}
\bibitem{Note} It should be noted that actually qRAM is not necessary for every machine learning algorithm. For example, principal component analysis. In such a case we only need multiple copies of a density matrix $\rho$ and the possibility to perform controlled swap operations. qRAM can be used to prepare the $\rho$s but so can other quantum subroutines.
\end{thebibliography}
\end{document}